\documentstyle[prl,aps,epsfig,twocolumn]{revtex}

\newcommand\be{\begin{eqnarray}}
\newcommand\ee{\end{eqnarray}}
\newcommand\nn{\nonumber}

\newcommand\ws{{\widetilde\sigma}}

\begin{document}
\title{Quantum memory for photons: I. Dark state polaritons}
\author{M. Fleischhauer$^1$ and M. D. Lukin$^2$}
\address{$^1$ Fachbereich Physik, Univ. Kaiserslautern, D-67663 Kaiserslautern,
Germany\\
$^2$ ITAMP, Harvard-Smithsonian Center for Astrophysics,
                Cambridge, MA~~02138, U.S.A.}
\date{\today}
\maketitle
\begin{abstract}
An ideal and reversible transfer technique  
for the quantum state between light and 
metastable collective states of matter is presented 
and analyzed in detail.
The method is based on the control of photon propagation 
in coherently driven 3-level atomic media, in 
which the group velocity is adiabatically reduced to zero. 
Form-stable 
coupled excitations of light and matter 
(``dark-state polaritons'')
associated with the propagation of quantum fields in 
Electromagnetically Induced Transparency are identified, their
basic properties discussed and their application for 
quantum memories for light analyzed.
\end{abstract}


\nobreak
                

\section{introduction}


Recent advances in quantum information science
lead to many interesting new concepts such as quantum computation, quantum
cryptography and teleportation \cite{q-comp,q-crypt,q-teleport}. 
The practical implementation of quantum processing protocols
requires coherent manipulation of a large 
number of coupled quantum 
systems which is an extremely difficult task.
One of the particular challenges for the implementation of these
ideas involves physically transporting or communicating quantum 
states between different 
nodes of quantum networks \cite{deVinzenco}. 
Quantum optical systems appear to be 
very attractive 
for the realization of such networks. On one hand 
photons are ideal carriers of quantum information:
they are fast, robust and readily available. On the other hand
atoms represent reliable and long-lived  storage and processing 
units. Therefore the challenge is to develop a technique for coherent 
transfer of  quantum information carried by light to atoms and vise versa.
In other words it is necessary to have a quantum
memory that is capable of storing and releasing quantum states on the
level of individual qubits and on demand. 
Such a device needs to be entirely coherent
and in order to achieve a unidirectional transfer (from field to atoms or
vise versa) an explicit time dependent control mechanism is required.  

{\it Classical} optical data storage in the time domain,
based on the phenomenon of spin-  \cite{Hahn55} and photon echo \cite{Abella66},
has a long history. After the first proposals of stimulated two-level photon 
echo \cite{MossbergOL82}
and demonstrations of light-pulse storage in these systems \cite{CarlsonOL83}
many important developments have taken place in this field. 
Particularly interesting are techniques based 
on Raman photon echos \cite{LeungOptCom82} as they
combine the long livetime of ground-state hyperfine or Zeeman coherences
for storage with data transfer by light at optical frequencies \cite{HemmerOL94}. 
While these techniques are very powerful for high-capacity storage of {\it classical}
optical data, they cannot be used for {\it quantum} memory purposes. The techniques
employ direct or dressed-state optical pumping and thus contain dissipative elements
or have other limitations in the transfer process between light and matter. 
As a consequence they do not operate on the level of individual photons and cannot be
applied for quantum information processing.

The conceptually simplest 
approach to a {\it quantum} memory for light 
is to ``store'' the state of a single 
photon in an individual atom. 
This approach  involves a coherent absorption and 
emission of single photons by single atoms. However, 
the single-atom absorption cross-section is very 
small, which makes such a process very inefficient.
A very elegant solution to this problem is provided
by cavity QED \cite{cavity-QED}. Placing an atom in 
a high-$Q$ 
resonator effectively enhances its cross-section by the number
of photon round-trips during the ring-down time and thus makes 
an effective transfer possible. Raman adiabatic passage techniques
\cite{STIRAP} 
with time-dependent external control fields can be used to implement
a directed but reversible transfer of the quantum state of a photon to
the atom (i.e. {\it coherent} absorption). 
However, despite the enormous experimental progress in this 
field \cite{cavity-QED-prob}, it is technically very challenging 
to achieve the necessary strong-coupling regime. 
Furthermore the single-atom system is by construction highly 
susceptible to the loss of atoms and the speed of operations is 
limited by the large $Q$-factor.

On the other hand a photon can be absorbed with unit probability 
in an optically thick ensemble of atoms. Normally such absorption is 
accompanied by  {\it dissipative} processes which result in decoherence 
and thus deteriorate the quantum state. 
Nevertheless it has been shown 
that such absorption of light leads to a 
partial mapping of its quantum properties
to atomic ensembles \cite{Polzik-old,Moelmer}. 
As a consequence of dissipation these methods do not allow 
to reversibly store the quantum state on the level of {\it individual} photon
wave-packets (single qubits). Rather a stationary 
source of identical copies 
is required (e.g. a stationary source of squeezed vacuum, which can be 
considered as a train of identical wave-packets in a squeezed vacuum state)
to partially map quantum statistics from light to matter.

Recently we have proposed a method that
combines the enhancement of the absorption cross section
in many-atom systems  with dissipation-free adiabatic passage techniques
\cite{Lukin00-ent,Fl00-pol,Fl00-OptCom}. 
It is based on an adiabatic 
transfer of the quantum state of photons to {\it collective atomic 
excitations} using electromagnetically induced transparency (EIT)
in 3-level atoms \cite{EIT}. Since the technique alleviates most of 
the stringent
requirements of single-atom cavity-QED, it could become the basis
for a fast and reliable quantum network. Recent experiments 
\cite{lui,phillips} have already 
demonstrated one of the basic principle of this technique - the dynamic 
group velocity reduction and adiabatic 
following in the so-called ``dark-state'' polaritons. 
The aim of the present and subsequent papers is to analyze the
physics of the reversible storage technique in detail and to discuss 
its potentials and limitations. 

Electromagnetically induced transparency can be used to 
make a resonant, opaque medium 
transparent by means of quantum interference. Associated with the
transparency is a large linear dispersion, which has been demonstrated 
to lead to a substantial reduction of the group velocity 
of light \cite{group}. 
Since the group velocity reduction is a linear process, the
quantum state of a slowed light pulse can be preserved. Therefore a
non-absorbing medium with a slow group velocity is in fact a temporary 
``storage'' device. However, such a system
has only limited ``storage'' capabilities. In particular the achievable
ratio of storage time to pulse length is limited by the 
square root of the medium opacity \cite{Hau99b}
and can practically attain only values on the order of 10 to 100. 
This limitation originates from the fact that a small group velocity
is associated with a narrow spectral acceptance window of 
EIT \cite{EIT-spectr} and  hence
larger delay times require larger initial pulse length.

The physics of the state-preserving slow light
propagation in EIT is associated with the existence of 
quasi-particles, which we call
dark-state polaritons (DSP). A dark-state polariton is a mixture of 
electromagnetic
and collective atomic excitations of spin
transitions (spin-wave). The mixing angle between the
two components determines the propagation velocity and 
is governed by the atomic density and the strength of 
an external control field. The key idea of the present approach 
is the dynamic rotation of the mixing angle which
leads to an adiabatic passage 
from a pure photon-like to a pure spin-wave polariton 
thereby decelerating the initial photon wavepacket 
to a full stop. In this process
the quantum state of the optical field 
is completely transferred to the atoms. During the
adiabatic slowing the spectrum of the pulse becomes narrower in 
proportion to the group velocity, which essentially eliminates the
limitations on initial spectral width or pulse length and 
very large ratios of storage time to initial pulse length can be achieved.
Reversing the rotation at a later time regenerates the photon wave-packet.
Hence the extension of EIT to a  dynamic group-velocity reduction 
via adiabatic following in polaritons can be used as the basis of an 
effective quantum memory. Before proceeding we note some earlier work on
the subject. The 
polariton picture of Raman adiabatic passage has first been introduced
in Ref. \cite{Mazets96}. Furthermore Grobe and coworkers
\cite{grobe} pointed out that 
the spatial profile of an atomic Raman coherence can be mirrored into
the electromagnetic field by coherent 
scattering, whereas time-varying fields can be used to 
create spatially non-homogeneous matter excitations. 

In the present paper we will present a quantum picture of slow light
propagation in EIT in terms of dark-state polaritons. We will analyze
the properties of the polaritons and discuss their application 
to reversible, fast
and high-fidelity quantum memories. Limitations and restrictions 
of the transfer process from non-adiabatic processes will be discussed
as well as effects from the medium boundary and atomic motion.
Other important aspects of the collective quantum memory such as 
its decoherence properties will be the subject of subsequent publications.


\section{quantum memory for a single-mode field}


The essential aspects of the quantum state 
mapping technique can be most easily
understood for the case of a single mode of the radiation field
as realized e.g. in a single mode optical cavity. In what follows
we will address this case first in order to motivate the following 
discussion on propagating photon wave packets.



\begin{figure}[ht]
\centerline{\epsfig{file=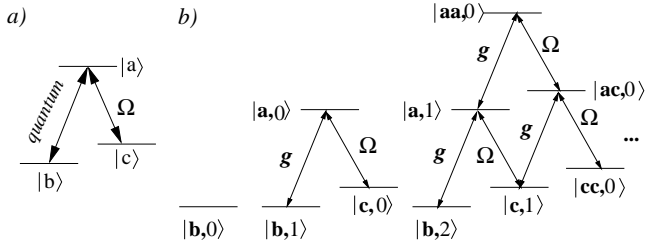,width=8.5 cm}}
\vspace*{2ex}
\caption{a) 3-level atoms coupled to single quantized mode and classical
control field of (real) Rabi-frequency $\Omega(t)$. b) coupling of relevant
bare eigenstates for at most one photon}
\label{3-level}
\end{figure}

Consider a collection of $N$ 3-level atoms 
with two meta-stable lower states as shown in Fig.~\ref{3-level} interacting 
with two single-mode optical fields. 
The transition $|a\rangle \to |b\rangle$ of each of these atoms is
coupled to a quantized radiation mode.
Moreover the transitions from $|a\rangle \to |c\rangle$ are resonantly driven
by a classical control field of Rabi-frequency $\Omega$. The dynamics of 
this system is described by the interaction Hamiltonian:
\begin{equation}
H = \hbar g \sum_{i = 1}^N  \hat a\sigma_{ab}^i + 
\hbar\Omega(t) {\rm e}^{-i\nu t}
\sum_{i = 1}^N  
\sigma_{ac}^i + {\rm h.c.} .  
\label{ham1}
\end{equation}
Here $\sigma_{\mu\nu}^i = |\mu\rangle_{ii}\langle \nu|$ is the 
flip operator of the $i$th atom between states $|\mu\rangle$ and 
$|\nu\rangle$.
$g$ is the coupling  constant between the atoms and the quantized field mode 
(vacuum Rabi-frequency) which for simplicity is assumed 
to be equal for all atoms.

When all atoms are prepared initially in  level $|b\rangle$ 
the only states coupled by the interaction are the totally symmetric
Dicke-like states \cite{Dicke54}
\be
|{\bf b}\rangle &=& |b_1,b_2,\dots,b_N\rangle,\\
|{\bf a}\rangle &=& \frac{1}{\sqrt{N}} \sum_{j=1}^N
|b_1,\dots,a_j,\dots,b_N\rangle,\\
|{\bf c}\rangle &=& \frac{1}{\sqrt{N}} \sum_{j=1}^N
|b_1,\dots,c_j,\dots,b_N\rangle,\\
|{\bf aa}\rangle &=& \frac{1}{\sqrt{2N(N-1)}}\sum_{i\ne j=1}^N
|b_1,\dots,a_i,\dots,a_j,\dots,b_N\rangle\\
&&{\rm etc.}\nonumber
\ee
In particular, if the field is initially in a state with at most one photon,
the relevant eigenstates of the bare system are: the total ground state
$|{\bf b},0\rangle$, which is not affected by the interaction at all,
the ground state with one photon in the field $|{\bf b},1\rangle$, as well
as the singly excited states $|{\bf a},0\rangle$ and $|{\bf c},0\rangle$.
For the case of two excitations, the interaction involves 3 more states etc.
The coupling of the singly and doubly excited 
systems are shown in  Fig.~\ref{3-level} b. 

The interaction has families of dark-states, i.e. states with zero adiabatic
eigenvalue \cite{cavity-QED,Lukin00-ent,Fl00-OptCom}. 
The simplest one is 
\be
|D,1\rangle &=& \cos\theta(t)\, |{\bf b},1\rangle -\sin\theta(t)\, 
|{\bf c},0\rangle,\label{dark1}\\
&&\tan\theta(t)=\frac{g\sqrt{N}}{\Omega(t)},\label{theta}
\ee
and in general one has
\be
&&|D,n\rangle=\nn\\
&&\enspace=\sum_{k=0}^n\sqrt{\frac{n!}{k!(n-k)!}}
(-\sin\theta)^k (\cos\theta)^{n-k}
|{\bf c}^k,n-k\rangle.\label{darkn}
\ee
The dark states do not contain the excited state and are thus immune to 
spontaneous emission. It should also 
be noted that although the dark states $|D,n\rangle$ are degenerate
they belong to exactly decoupled subsystems as long as spontaneous emission is
disregarded. This means there is no transition between them even if 
non-adiabatic corrections are taken into account.
The existence of collective dark states provides a very elegant way
to transfer the quantum state of the single-mode field to collective atomic 
excitations. Adiabatically rotating the mixing angle $\theta$ from $0$ to 
$\pi/2$ leads to a complete and reversible transfer of the 
photonic state to a collective atomic
state if the total number of excitations $n$ is less than the number
of atoms. 
This can be seen very easily from the expression for the dark states,
eq.(\ref{darkn}): If $\theta: 0\to\pi/2$ one has for all $n\le N$
\be
|D,n\rangle :\, |{\bf b}\rangle|n\rangle 
\longrightarrow |{\bf c}^n\rangle |0\rangle.
\ee
Thus if the initial quantum state of the single-mode light field 
is in any mixed state described by a density matrix 
$\hat\rho_f=\sum_{n,m} \rho_{nm}\, |n\rangle\langle m|$, the 
transfer process generates a quantum state of collective excitations
according to
\be
&&\sum_{n,m}\rho_{nm} \, |n\rangle\langle m|
 \otimes |{\bf b}\rangle\langle{\bf b}|
\longrightarrow\nn\\
&&\qquad\qquad |0\rangle\langle 0|\otimes 
\sum_{n,m} \rho_{nm}\, |{\bf c}^n\rangle\langle {\bf c}^m|.
\ee
It should be noted that the quantum-state transfer 
does not necessarily constitute a
transfer of energy from the quantum field to the atomic ensemble. 
Since in the Raman process the coherent ``absorption''
of a photon from the quantized mode is followed by a 
stimulated emission into the classical control field, most of the energy
is actually deposited in the latter field. 

The transfer of quantum states between light and matter due to adiabatic 
following in collective dark states is the key point of the present work. 
Before proceeding we also note that the  transfer rate is proportional to 
the total number of atoms $N$, which is a signature of collective coupling. 
This makes the proposed method potentially fast and robust.


\section{quantum description of slow-light propagation}



We now discuss a generalization of the mapping technique to propagating 
fields. 
The adiabatic transfer of the quantum state from the radiation mode to 
collective atomic excitations discussed in the previous section
is strongly related to intracavity electromagnetically induced transparency
(EIT) \cite{LukinOL_98}.
In order to generalize the technique
to multi-mode fields it is useful to discuss first 
the propagation of light in 3-level media under conditions of EIT.


\subsection{model}

\begin{figure}[ht]
\centerline{\epsfig{file=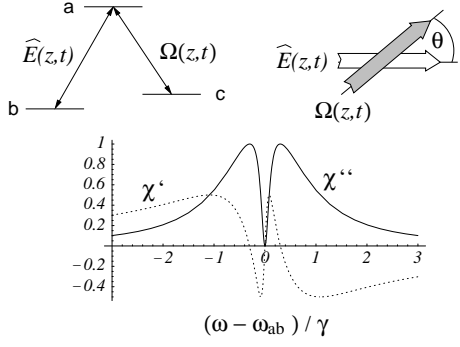,width=6.0cm}}
\vspace*{2ex}
\caption{{\it top:} 3-level $\Lambda$-type medium resonantly coupled to a classical
field with Rabi-frequency $\Omega(t)$ and quantum field $\hat E(z,t)$.
{\it bottom:} Typical susceptibility spectrum for probe field $\hat E$ 
as function of normalized detuning from resonance
for resonant drive field. Real part $\chi^\prime$ describes refractive index
contribution and imaginary part $\chi^{\prime\prime}$ absorption.
}
\label{1-d}
\end{figure}

Consider the quasi 1-dimensional problem 
shown in Fig.~\ref{1-d}. A quantized
electromagnetic field with positive frequency part of 
the electric component $\hat E^{(+)}$
couples resonantly
the transition between the ground state $|b\rangle$ and
the excited state $|a\rangle$. $\nu=\omega_{ab}$ is the carrier frequency 
of the optical field.
The upper level $|a\rangle$ is furthermore coupled to the stable state 
$|c\rangle$ via a coherent control field
with Rabi-frequency $\Omega$.


The interaction Hamiltonian reads
\be
\hat V&=&-\wp \sum_j\Bigl(\hat\sigma_{ab}^j\, \hat E^{(+)}(z_j)+h.a\Bigr)\nn\\
&& -\hbar \sum_j\Bigl(\hat\sigma_{ac}^j\, \Omega(z_j,t) \, {\rm e}^{i 
(k_d^\parallel z_j 
-\nu_d t)}
   +h.a\Bigr)
\ee
where $z_j$ denotes the position of the $j$th atom, $\wp$ denotes the
dipole matrix element between the states $|a\rangle$ and $|b\rangle$, and
\be
\hat\sigma_{\alpha\beta}^j \equiv |\alpha_j\rangle \langle \beta_j|
\ee
defines the atomic flip operators. $k_d^\parallel = 
\vec k_d \cdot \vec {\rm e}_z
=\frac{\nu_d}{c}\cos\vartheta$ is the projection of the wavevector of the
control field to the propagation axis of the quantum field.
For the sake of simplicity we here 
assume that the carrier frequencies
$\nu$ and $\nu_d$ of the quantum and control fields coincide with the
atomic resonances $\omega_{ab}$ and $\omega_{ac}$ respectively. 
Motional effects and the associated Doppler-shifts will be discussed later.
We introduce slowly-varying variables according to
\be
\hat E^{(+)}(z,t) &=& \sqrt{\frac{\hbar \nu }{2\varepsilon_0 V}}
\hat{\cal E}(z,t)\, {\rm e}^{i\frac{\nu}{c}(z-ct)},\\
\hat\sigma_{\mu\nu}^j(t) &=& {\widetilde\sigma}_{\mu\nu}^j(t)
\, {\rm e}^{-i\frac{\omega_{\mu\nu}}{c}(z-ct)}.
\ee
Here $V$ is some quantization volume, which for simplicity 
was chosen to be equal to the interaction volume.

If the (slowly-varying) quantum amplitude does not change in a length
interval $\Delta z$ which contains $N_z\gg 1$ atoms we can introduce 
continuum atomic variables
\be
{\widetilde\sigma}_{\mu\nu}(z,t) =
\frac{1}{N_z}\sum_{z_j\in N_z}\, {\widetilde\sigma}_{\mu\nu}^j(t)
\ee
and make the replacement $\sum_{j=1}^N 
\longrightarrow \frac{N}{L}\int {\rm d} z $, where $N$ 
is the number of atoms, and 
$L$ the length of the interaction volume in propagation direction
of the quantized field.
This yields the continuous form of the interaction Hamiltonian
\be
{\hat V} &=& -
\int\!\! \frac{{\rm d} z}{L} 
\biggl( \hbar g N 
{\widetilde\sigma}_{ab}(z,t)\, \hat{\cal E}(z,t)\nn\\
&&\qquad\qquad
+\hbar \Omega(z,t){\rm e}^{i\Delta k z}
N(z){\widetilde\sigma}_{ac}(z,t)+h.a.\biggr).
\label{ham}
\ee
Here $g=\wp\sqrt{\frac{\nu}{2\hbar\epsilon_0 V}}$ is the atom-field 
coupling constant and $\Delta k= k_d^\parallel-k_d=\frac{\omega_{ac}}{c}
\left(\cos\vartheta-1\right)$.

The evolution of the Heisenberg operator corresponding to the 
quantum field can be described in slowly varying 
amplitude approximation by the propagation equation
\begin{eqnarray}
\left(\frac{\partial}{\partial t}+c\frac{\partial}{\partial z}\right)
\hat {\cal E}(z,t)= { i} g N\, \ws_{ba}(z,t).\label{field}
\end{eqnarray}
The atomic evolution is governed by a set of 
Heisenberg-Langevin equations
\begin{eqnarray}
\dot{\widetilde\sigma}_{aa} &=&-\gamma_a {\widetilde\sigma}_{aa}
-ig\Bigl(\hat{\cal E}^\dagger{\widetilde\sigma}_{ba}-h.a.\Bigr)\nn\\
&&-i\Bigl(\Omega^*{\rm e}^{-i\Delta k z}{\widetilde\sigma}_{ca}-h.a\Bigr)+F_a
,\\
\dot{\widetilde\sigma}_{bb} &=&\gamma {\widetilde\sigma}_{aa}
+ig\Bigl(\hat{\cal E}^\dagger{\widetilde\sigma}_{ba}-h.a.\Bigr)+F_b,\\
\dot{\widetilde\sigma}_{cc} &=&\gamma^\prime {\widetilde\sigma}_{aa}
+i\Bigl(\Omega^*{\rm e}^{-i\Delta k z}{\widetilde\sigma}_{ca}-h.a\Bigr)
+F_c,\\
\dot{\widetilde\sigma}_{ba} &=&-\gamma_{ba} {\widetilde\sigma}_{ba}
+ig{\cal E}\Bigl({\widetilde\sigma}_{bb}-{\widetilde\sigma}_{aa}\Bigr)\nn\\
&&
+i\Omega{\rm e}^{i\Delta k z}{\widetilde\sigma}_{bc} +F_{ba},\\
\dot{\widetilde\sigma}_{ca} &=&-\gamma_{ca} {\widetilde\sigma}_{ca}
+i\Omega{\rm e}^{i\Delta k z}
\Bigl({\widetilde\sigma}_{cc}-{\widetilde\sigma}_{aa}\Bigr)\nn\\
&&
+ig\hat{\cal E}{\widetilde\sigma}_{ba} +F_{ca},\\
\dot{\widetilde\sigma}_{bc} &=&
i\Omega^*{\rm e}^{-i\Delta k z}{\widetilde\sigma}_{ba}
-ig\hat {\cal E}{\widetilde\sigma}_{ac},\label{sigma_bc_eq}
\end{eqnarray}
$\gamma_a=\gamma+\gamma^\prime$ and
$\gamma, \gamma^\prime$ denote longitudinal
and 
$\gamma_{\mu\nu}$ transversal decay rates.  $F_\mu$ and $F_{\mu\nu}$
are $\delta$-correlated Langevin noise operators, whose
explicit form is not of interest here. 

It should be noted that we have
disregarded dissipative population exchange processes due to e.g.
spin-flip collisions and dephasing of the lower-level transition.
This is justified since we assume that the 
interaction time is sufficiently short
compared to the characteristic times of these processes.
Both, dissipative population exchange as well
as dephasing will be discussed in detail later on.


\subsection{low-intensity approximation}


In order to solve the propagation problem, 
we now assume that the Rabi-frequency of the quantum field
is much smaller than $\Omega$ and that the
number density of photons in the input pulse is much less than the
number density of atoms. 
In such a case the atomic 
equations can be treated perturbatively in $\hat {\cal E}$. 
In zeroth order only ${\widetilde\sigma}_{bb}={\bf 1}$ is different from zero
and in first order 
one finds 
\begin{eqnarray}
\ws_{ba}=-\frac{ i}{\Omega^*}  {\rm e}^{i\Delta k z}
\frac{\partial}{\partial t}
\ws_{bc}. \label{bc}
\end{eqnarray}
With this the interaction of the probe pulse with the medium 
can be described by the amplitude of the probe electric field
$\hat {\cal E}$ and the collective ground-state spin variable $\ws_{bc}$:
\be 
\left(\frac{\partial}{\partial t}+c\frac{\partial}{\partial z}\right)
\hat {\cal E}(z,t) = \frac{g  N}{\Omega^*}{\rm e}^{i\Delta k z}
\frac{\partial}{\partial t}\ws_{bc},\label{E_lin}
\ee
and
\be
&&\ws_{bc}=
-\frac{g \hat {\cal E}}{\Omega} {\rm e}^{-i\Delta k z}\nn\\
&&\enspace -\frac{ i}{\Omega}\left[
\left(\frac{\partial}{\partial t}+\gamma_{ba}\right)\left(-\frac{i}{\Omega^*}
\frac{\partial}{\partial t}\ws_{bc}\right)+{\rm e}^{-i\Delta k z}
 F_{ba}\right].
\label{s_cb_2}
\ee
%
%


\subsection{adiabatic limit}


The propagation equations simplify considerably if we assume
a sufficiently slow change of $\Omega$, i.e. adiabatic conditions
\cite{Fl-Manka96,Hau99b,LukinNLO}.
Normalizing the time to a characteristic scale $T$ via
$\tilde t=t/T$
and expanding the r.h.s. of (\ref{s_cb_2}) in powers of $1/T$
we find in lowest non-vanishing order  
\be
\ws_{bc}(z,t)
=-g\frac{\hat{\cal E}}{\Omega}{\rm e}^{-i\Delta k z}
.\label{s_cb_ad}
\ee
We note that also the noise operator
$F_{ba}$ gives no contribution in
the adiabatic limit, since
$\langle  F_x(t)  F_y(t')\rangle \sim \delta(t-t')
= \delta(\tilde t-\tilde t')/T$.
Thus in the perturbative and adiabatic limit
the propagation of the quantum light pulse is governed by
\begin{eqnarray}
\left(\frac{\partial}{\partial t}+c\frac{\partial}{\partial z}\right)
\hat {\cal E}(z,t)&=& -\frac{g^2 N}{
\Omega^*}
 \frac{\partial}{\partial t}
\frac{\hat {\cal E}(z,t)}{\Omega}\label{field_ad}.
\end{eqnarray}
%
%


\subsection{slow light and delay-time limitations}


If $\Omega(z,t)=\Omega(z)$ is {\it constant in time}, the term on the r.h.s. 
of the propagation equation (\ref{field_ad})
simply leads to a modification
of the group velocity of the quantum field according to 
\be
v_g&=&v_g(z)=\frac{c}{1+n_g(z)},\\
n_g(z)&=&\frac{g^2 N}{|\Omega(z)|^2}
=\frac{3}{8\pi^2} \rho\lambda^3 \frac{kc\, \gamma}{|\Omega(z)|^2},
\ee
with $\rho$ being the atom density and $\lambda$ the resonant 
wavelength of the $a\to b$ transition.
The solution of the wave equation is in this case
\be
\hat{\cal E}(z,t)=\hat{\cal E}
\biggl(0,t-\int^z_0{\rm d}z'\, \frac{1}{v_g(z')}\biggr),\label{sol-z}
\ee
where $\hat{\cal E}(0,t')$ denotes the field entering the
interaction region at $z=0$. 
This solution describes a propagation with a 
spatially varying velocity $v_g$.
It is apparent that the {\it temporal} profile of the pulse is
unaffected by the slow-down. As a consequence the integrated 
electromagnetic energy flux through 
a plane perpendicular to the propagation is the same at any position.
Furthermore the spectrum of the pulse 
remains unchanged
\be
S(z,\omega)&\equiv &\int_{-\infty}^\infty \!\!{\rm d}\tau\,
{\rm e}^{-i\omega\tau}\,
\Bigl\langle \hat{\cal E}^\dagger(z,t)\hat{\cal E}(z,t-\tau)\Bigr\rangle\nn\\
&=& S(0,\omega).
\ee
In particular the spectral width stays constant
\be
\Delta\omega_p(z)=\Delta\omega_p(0).
\ee
On the other hand a spatial change of the group velocity, either in a
stepwise fashion as e.g. at the entrance of the medium or in a 
continuous way e.g. due to a spatially decreasing control field,
leads to a {\it compression of the spatial pulse profile}. In particular, 
if the group velocity is {\it statically} reduced to a value $v_g$, 
the spatial pulse length $\Delta l$ is modified according to 
\be
\Delta l=\frac{v_g}{c}\, \Delta l_0,\label{deltaz},
\ee
as compared with a free-space value $\Delta l_0$. 

It is instructive to discuss the limitations to the
achievable delay (and therefore ``storage'') time $\tau_d$
in an EIT medium with a very small but finite group
velocity. A first and obvious limitation is set by the
finite lifetime of the dark state, which had been neglected
in the previous subsection. If $\gamma_{bc}$ denotes the dephasing
rate of the $b-c$ transition it is required that $\tau_d\le \gamma_{bc}^{-1}$.
A much stronger limitation arises however from the violation of the adiabatic
approximation.

The EIT medium behaves like a non-absorbing, linear 
dispersive medium only within a certain frequency window around the
two-photon resonance (see Fig.\ref{1-d} and \cite{EIT-spectr}). 
The adiabatic approximation made in the last sub-section essentially 
assumes that everything happens 
within this frequency window. If the pulse becomes too short, 
or its spectrum 
too broad relative to the transparency width, absorption 
and higher-order dispersion need to be 
taken into account.

The transparency window is defined by the intensity transmission of the
medium. 
In order to determine its width we consider 
the susceptibility of an ideal, homogeneous 
EIT medium with a resonant drive field. Here one has
\be
\chi&=&\frac{n_g}{kc} \frac{|\Omega|^2 \delta}
{|\Omega|^2 -\delta^2 -i\gamma \delta}\nn\\
&\approx & \frac{n_g}{kc}\left[\delta +i\delta^2 \frac{\gamma}{
|\Omega|^2} +{\cal O}(\delta^3)\right],
\ee
where $\delta=\nu-\omega_{ac}$ is the detuning of the probe field.
If we assume a homogeneous drive field, 
we find
\be
T(\delta,z)&=&\exp\Bigl\{-k z \,{\rm Im}[\chi]\Bigr\}
\nn\\
&\approx&\exp\Bigl\{-\delta^2/\Delta\omega_{tr}^2\Bigr\},
\ee
with
\be
\Delta\omega_{tr}=\left[\frac{c}{\gamma l}
\frac{|\Omega|^2}{n_g}\right]^{1/2}=\frac{|\Omega|^2}{\gamma}
\frac{1}{\sqrt{\alpha}},
\ee
$l$ being the propagation length in the medium and 
$\alpha\equiv 
\frac{3}{8\pi^2}\rho\lambda^3
k l$ the opacity in the absence of EIT.
One recognizes that the transparency width decreases with increasing
group index.  
It is instructive to express the transparency width in terms of the
pulse delay time $\tau_d=n_g l/c$ for the medium. This yields
\be
\Delta\omega_{tr}=\sqrt{\alpha}\, \frac{1}{\tau_d}.
\ee
Hence large delay times imply a narrow transparency window, which in turn 
requires a long pulse time. 
%
\noindent When the group velocity gets too small such that 
the transparency window becomes smaller than the spectral width of the
pulse, the adiabatic condition is violated, and the pulse is absorbed. 
Hence there is an upper bound for the ratio of achievable delay (storage)
time to the initial pulse length of a photon wavepacket
\be
\frac{\tau_d}{\tau_p} \le \sqrt{\alpha}\quad{\rm and}\quad 
\tau_d\le \gamma_{bc}^{-1}.
\ee
The ratio $\tau_d/\tau_p$ is the figure of merit for a memory 
device. The larger this ratio the better suited is the system
for a temporal storage. 

There is another quantity which is important
for the storage capacity, namely the ratio of medium length $L$ to the
length $L_p$ 
of an individual pulse inside the medium.  Following similar arguments
as above one finds that this quantity is also limited by the square root
of the opacity
\be 
\frac{L}{L_p}\le \sqrt{\alpha},
\ee
where $n_g\gg 1$ was assumed.

In practice, the achievable opacity $\alpha$ of atomic vapor systems is limited
to values below $10^4$ resulting in upper bounds for the 
ratio of time delay to pulse length of the order of 100.
Thus dense EIT media with ultra-small group velocity are only of 
limited use as a temporary storage device. 
The propagation velocity cannot be made zero and 
non-adiabatic corrections 
limit the achievable ratio
of storage time to the time length of an individual qubit. 

It should be mentioned that the narrowing of the EIT transparency window 
is also a consideration for effects involving freezing light in a 
moving media  \cite{Olga-slow} and so-called optical black holes based on EIT 
\cite{Leonhardt00}. In the following we will show, that EIT can nevertheless
be used for memory purposes in a very effective way when combined with
adiabatic passage techniques.


\section{dark-state polaritons}


In the previous section we have discussed the propagation of a quantum field
in an EIT medium under otherwise stationary conditions, i.e. 
with a constant or only spatially varying control field.
Since under these conditions, the hamiltonian of the system is 
{\it time-independent}, a coherent process that allows for 
an uni-directional transfer
of the quantum state of a photon wavepacket to the atomic ensemble
was not possible. We will show now that this limitation can be overcome
easily by allowing for a {\it time-dependent} control field.
This provides 
an elegant tool to control the propagation of a quantum light pulse.
For a spatially homogeneous but time-dependent control field,
$\Omega=\Omega(t)$, the
propagation problem can be solved in a very instructive way in 
a quasi-particle picture. In the following we will introduce these
quasi-particles, called dark-state polaritons \cite{Fl00-pol,Mazets96}
and discuss their properties, applications and limitations.


\subsection{definition of dark- and bright-state polaritons}


Let us consider the case of a time-dependent, spatially homogeneous
and real control field  $\Omega=\Omega(t)=\Omega(t)^*$. 
We introduce a rotation in the space of physically relevant variables
--  the electric field $\hat {\cal E}$ and the atomic spin 
coherence $\ws_{bc}$ -- 
defining two new quantum fields 
$\hat \Psi(z,t)$ and $\hat \Phi(z,t)$
\be
\hat\Psi &=&
\cos\theta(t)\, \hat {\cal E}(z,t) - \sin\theta(t)\, \sqrt{N}\,
\ws_{bc}(z,t)\, {\rm e}^{i\Delta k z},\\
\hat\Phi &=&
\sin\theta(t)\, \hat {\cal E}(z,t) + \cos\theta(t)\, \sqrt{N}\,
\ws_{bc}(z,t)\, {\rm e}^{i\Delta k z},
\ee
with the mixing angle $\theta(t)$ given by
\be
\tan^2\theta(t) = \frac{g^2 N}{\Omega^2(t)}=n_{\rm g}(t).
\ee
$\hat \Psi$ and $\hat\Phi$ are superpositions 
of electromagnetic ($\hat{\cal E}$) 
and collective atomic  components ($\sqrt{N}\ws_{bc}$), whose
admixture can be controlled through $\theta(t)$
by changing the strength of the external driving field.

Introducing a plain-wave decomposition 
$\hat\Psi(z,t)=\sum_k \hat\Psi_k(t)\, {\rm e}^{ikz}$ 
and $\hat\Phi(z,t)=\sum_k\hat\Phi_k(t)\, {\rm e}^{ikz}$ respectively,
one finds that
the mode operators obey the commutation
relations
\begin{eqnarray}
\Bigl[\hat\Psi_k, \hat\Psi_{k'}^+\Bigr] &=&
 \delta_{k,k'}\, \Bigl[\cos^2\theta + 
\sin^2\theta\frac{1}{N}\sum_j
({\hat \sigma}_{bb}^j-{\hat \sigma}_{cc}^j)\Bigr],\\
\Bigl[\hat\Phi_k, \hat\Phi_{k'}^+\Bigr] &=&
 \delta_{k,k'}\, \Bigl[\sin^2\theta + 
\cos^2\theta\frac{1}{N}\sum_j
({\hat \sigma}_{bb}^j-{\hat \sigma}_{cc}^j)\Bigr],\\
\Bigl[\hat\Psi_k, \hat\Phi_{k'}^+\Bigr] &=&
 \delta_{k,k'}\sin\theta\cos\theta\, \Bigl[1-\frac{1}{N}\sum_j
({\hat \sigma}_{bb}^j-{\hat \sigma}_{cc}^j)\Bigr].
\end{eqnarray}
In the linear limit considered here, where the number density
of photons is much smaller than the density of atoms,
${\hat \sigma}_{bb}^j \approx 1, {\hat \sigma}_{cc}^j \approx 0$.
Thus the new fields possess bosonic commutation relations 
\be
\Bigl[\hat\Psi_k, \hat\Psi_{k'}^+\Bigr] \approx 
\Bigl[\hat\Phi_k, \hat\Phi_{k'}^+\Bigr] &\approx & \delta_{k,k'},\\
\Bigl[\hat\Psi_k, \hat\Phi_{k'}^+\Bigr] &\approx & 0,
\ee
and we can
associate with them bosonic quasi-particles (polaritons). 
Furthermore one immediately verifies that all number states created by
$\hat\Psi_k^\dagger$ 
\begin{eqnarray}
|n_k\rangle = \frac{1}{\sqrt{n!}}
\Bigl(\hat\Psi_k^\dagger\Bigr)^n |0\rangle |b_1 ...b_N\rangle,
\label{dark} 
\end{eqnarray}
where $|0\rangle$ denotes the field vacuum,  
are dark-states \cite{dark,Lukin00-ent}.
The states $|n_k\rangle$ do not contain the excited
atomic state and are thus immune to spontaneous emission. Moreover, 
they are eigenstates of the interaction Hamiltonian with eigenvalue zero, 
\be
{\hat V}\, |n_k\rangle  = 0.
\ee
For these reasons we call the quasi-particles $\hat \Psi$
``dark-state polaritons''.
Similarly one finds that the elementary excitations of $\hat \Phi$
correspond to the bright-states in 3-level systems. Consequently
these quasi-particles are called ``bright-state polaritons''.

One can transform the equations of motion for the electric field and
the atomic variables into the new field variables. In the 
low-intensity approximation one finds
\be
\biggl[\frac{\partial}{\partial t} +c\cos^2\theta
\frac{\partial}{\partial z}\biggr]\,
\hat\Psi =-\dot\theta\, \hat\Phi -\sin\theta
\cos\theta\, c\frac{\partial}{\partial z}\hat\Phi\label{Psi-full}
\ee
and 
\be 
\hat\Phi &=&\frac{\sin\theta}{g^2 N}
\biggl(\frac{\partial}{\partial t}+\gamma\Bigr)\Bigl(\tan\theta 
\frac{\partial}{\partial t}\biggr)
\Bigl(\sin\theta\,\hat\Psi-\cos\theta\, \hat\Phi\Bigr)\label{Phi-full}\\
&& +i \frac{\sin\theta}{g\sqrt{N}} F_{ba},\nonumber
\ee
where one has to keep in mind that the mixing angle $\theta$ 
is a function of time.


\subsection{adiabatic limit}


Introducing the adiabaticity 
parameter $\varepsilon\equiv \Bigl(g\sqrt{N} 
T\Bigr)^{-1}$ with $T$ being a characteristic time, one can 
expand the equations of motion in powers of $\varepsilon$.
In lowest order i.e. in the
adiabatic limit one finds
\be
\hat\Phi &\approx & 0.
\ee
Consequently
\be
\hat{\cal E}(z,t) &=& \enspace\cos\theta(t)\, \hat\Psi(z,t),\label{E-Psi}\\
\sqrt{N} \ws_{bc} &=& -\sin\theta(t)\, \hat\Psi(z,t)\, {\rm e}^{-i\Delta k z}.
\label{sigma-Psi}
\ee
Furthermore
$\hat\Psi$ obeys the very simple equation of motion
\begin{eqnarray}
\left[\frac{\partial}{\partial t}+c\cos^2\theta(t)
\frac{\partial}{\partial z}\right]\hat\Psi(z,t)=0.\label{Psi-eq}
\end{eqnarray}
%
%


\subsection{``stopping'' and re-accelerating photon wavepackets}


Eq.(\ref{Psi-eq}) describes a shape- and quantum-state 
preserving propagation with 
instantaneous velocity
$v=v_g(t)=c\cos^2\theta(t)$: 
\begin{equation}
\hat\Psi(z,t)=\hat \Psi\biggl(z- c\int^t_0\!\!\!{\rm d}\tau
\cos^2\theta(\tau),0\biggr).
\label{sol}
\end{equation}
For $\theta\to 0$, i.e. for a strong external drive field $\Omega^2\gg g^2 N$,
the polariton has purely photonic character $\hat\Psi =\hat{\cal E}$
and the propagation velocity is that of the vacuum speed of light.
In the opposite limit of a weak drive field $\Omega^2\ll g^2 N$ such that
$\theta\to \pi/2$, the polariton becomes spin-wave like
$\hat\Psi =-\sqrt{N}\ws_{bc} {\rm e}^{i\Delta k z}$ and its
propagation velocity approaches zero. Thus the following mapping can be 
realized
\be
\hat{\cal E}(z) \, \Longleftrightarrow \, \widetilde\sigma_{bc}(z') {\rm e}^
{i\Delta k z'}
\ee
with $z'=z+z_0=z+\int^\infty_0\!\!{\rm d}\tau c\cos^2\theta(\tau)$. 
This is the essence of the transfer technique of quantum states from 
photon wave-packets propagating at the speed of light
to stationary atomic excitations 
(stationary spin waves). Adiabatically rotating the mixing angle from
$\theta=0$ to $\theta=\pi/2$ decelerates the polariton to a full stop,
changing its character from purely electromagnetic to purely atomic.
Due to the linearity of eq.(\ref{Psi-eq}) and the conservation of
the spatial shape, the quantum state of the polariton is not changed
during this process. 

\begin{figure}[ht]
\centerline{\epsfig{file=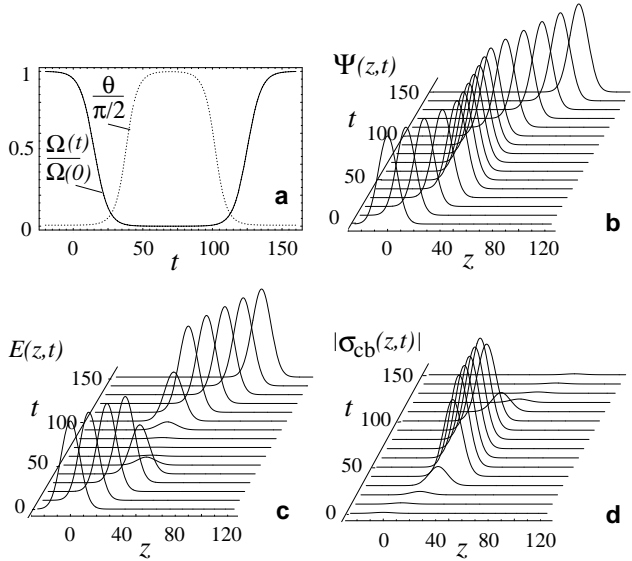,width=9.0cm}}
 \vspace*{2ex}
 \caption{Propagation of a dark-state polariton 
with envelope $\exp\{-(z/10)^2\}$. 
The mixing angle is rotated from $0$ to $\pi/2$ and back
according to $\cot\theta(t)  =$ $ 100(1-0.5 \tanh[0.1(t-15)]$ $
 + 0.5\tanh[0.1(t-125)])$ as shown in (a). 
 The  
coherent amplitude of the polariton $\Psi=\langle\hat\Psi\rangle$ 
is plotted in (b) and the electric field $E=\langle \hat E\rangle
$ and matter components $|\sigma_{cb}|=|\langle\hat \sigma_{cb}\rangle|$
in (c) and (d) respectively. 
 Axes are in arbitrary units with $c=1$. }
\label{stop}
\end{figure}

Likewise the polariton can be
re-accelerated to the vacuum speed of light; in this process the stored 
quantum state is transferred back to the field. 
This is illustrated in Fig.\ref{stop}, where we have shown 
the coherent amplitude of a dark-state polariton which results from an initial
light pulse as well as the corresponding  field and matter
components. 

\subsection{simultaneous narrowing of transparency window and pulse
spectral width}


As was discussed above, the transparency window
of the EIT medium, i.e. the range of frequencies for which absorption
is negligible, decreases with the group velocity. Since 
the bandwidth of the propagating  pulse should always be 
contained within this range of frequencies to avoid absorption, the 
question arises whether this prevents the stopping of the polariton.

\begin{figure}[ht]
\centerline{\epsfig{file=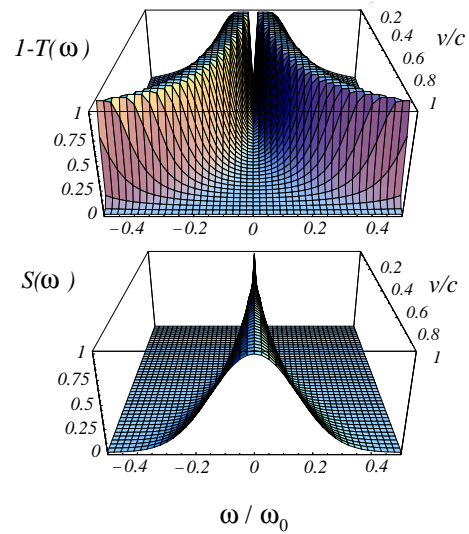,width=6 cm}}
\vspace*{2ex}
\caption{Simultaneous narrowing of transmission spectrum (top) and pulse 
spectrum (bottom) for time dependent variation of group velocity
$v/c$ in units of $\omega_0= g^2 N/\gamma$. Parameters are
 $\alpha=20$, $g^2 N/\gamma^2=10$.}
\label{parallel_narrowing}
\end{figure}

To answer this question we first note that during the process of 
adiabatic slowing  
the {\it spatial} profile, and in particular the 
length of the wavepacket 
($\Delta l$) remains unaffected, as long as 
the group velocity is only a function of $t$.
I.e.
\be
\Delta l =\Delta l_0.
\ee
At the same time, the amplitude of the electric field
gets reduced and its temporal profile stretched 
due to the reduction of the group velocity. 
The opposite happens when the group velocity is increased.
One finds from eq.(\ref{E-Psi})
\be
\hat{\cal E}(z,t)=\frac{\cos\theta(t)}{\cos\theta(0)}\, \hat{\cal E}\biggl(z-
c\int^t_0\!\!\!{\rm d}\tau
\cos^2\theta(\tau),0\biggr).
\ee
As a consequence the spectrum of the probe field
changes during propagation. Assuming that $\cos\theta$ changes only
slow compared to the field amplitude one finds
\be
S(z,\omega) &=& \frac{\cos^2\theta(t)}{\cos^2\theta(0)}
 S\left(0,\frac{\omega}{\cos^2\theta(t)}\right).
\ee
In particular the spectral width narrows (broadens) according to 
\be
\Delta\omega_p(t)\approx \Delta\omega_p(0) \, 
\frac{\cos^2\theta(t)}{\cos^2\theta(0)}.
\ee
When the group velocity of the polariton is reduced in time due to
reduction of the control-field amplitude  
the EIT transparency window shrinks according to
\be
\Delta\omega_{tr}(t)=\frac{\cot^2\theta(t)}{\cot^2\theta(0)} \Delta\omega_{tr}(0).
\ee
 However the
spectral width of the wavepacket shrinks as well and the
ratio remains finite:
\be
\frac{\Delta\omega_{p}(t)}{\Delta\omega_{tr}(t)}
=\frac{\sin^2\theta(t)}{\sin^2\theta(0)}\frac{\Delta\omega_{p}(0)}
{\Delta\omega_{tr}(0)}.
\ee
As will be discussed later on, for practically relevant cases, 
$\sin^2\theta(t)/\sin^2\theta(0)$
is always close to unity.
Thus absorption can be prevented in the dynamic light-trapping method as long
as the input pulse spectrum lies within the {\it initial} transparency window:

\be
\Delta\omega_{p}(0)\ll \Delta\omega_{tr}(0).
\label{Delta-omega0}
\ee
As we have already noted earlier this condition can easily be fulfilled if an 
optically dense  medium is used. The simultaneous reduction in transparency 
bandwidth and pulse bandwidth is illustrated in Fig.\ref{parallel_narrowing}.  

\subsection{boundary behavior and initial pulse compression}


In the above discussion we have analyzed the propagation of the
probe pulse inside the medium. We now turn to the behavior at the
medium boundaries. There are two issues of interest: (i)  reflection from
the medium surface and (ii) the effects of a possible 
steplike change of the group velocity when passing from vacuum or air 
to the medium. 

\begin{figure}[ht]
\centerline{\epsfig{file=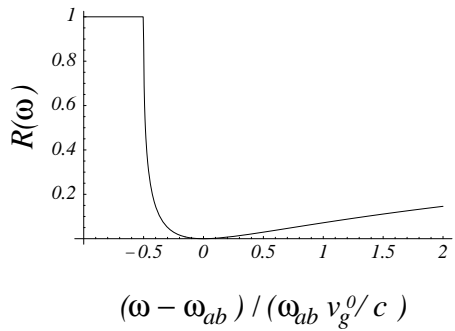,width=6 cm}}
\vspace*{2ex}
\caption{Reflection coefficient for normal incidence as function of
probe detuning from resonance.}
\label{boundary}
\end{figure}

Under ideal conditions the refractive index of the EIT medium is exactly
unity on resonance and hence there is no reflection for the resonant 
component of the input pulse.
If the spectrum of the input pulse is furthermore 
sufficiently narrow as compared 
to the initial transparency window of EIT, the refractive index is 
very close to unity 
over the entire relevant bandwidth and no field component gets reflected. 
To quantify this we note, that the index of refraction  near resonance of an
idealized, resonantly driven 3-level medium can be written as
\be
n(\omega)\approx \sqrt{
1 + \frac{2c}{v_g^0} \frac{(\omega-\omega_{ab})}{\omega_{ab}}}.
\ee
The reflection coefficient for normal incidence from vacuum to
the medium given by
\be
R(\omega)=\left|\frac{1-n(\omega)}{1+n(\omega)}\right|^2
\ee
is plotted in fig.\ref{boundary}.


\noindent Near resonance $R(\omega)$ 
can be approximated by 
\be
R(\omega)\approx \frac{\Delta\omega^2}{\Bigl(2\frac{v_{\rm g}^0}{c}
\omega_{ab} + \Delta\omega\Bigl)^2}
\ee
where $\Delta\omega =\omega-\omega_{ab}$. 
Since typically $\Delta\omega_p(0) \ll \omega_{ab} \,  v_{\rm g}^0/c$
reflection can be neglected.

When the group velocity inside the medium is smaller than c at the
time when the probe pulse enters the medium, the pulse
gets spatially compressed according to 
eq.(\ref{deltaz}). This  substantially reduces 
the requirements on the size of the medium and 
enhances the storage capacity.

Continuity of the electric field amplitude ${\hat{\cal E}}$ at the
entrance of the medium ($z=0$) implies a jump in the
polariton amplitude 
\be
\hat\Psi(+0,t)=\sqrt{\frac{c}{v_{\rm g}^0}}\, \hat\Psi(-0,t) =
\sqrt{\frac{c}{v_{\rm g}^0}}\,
\, \hat{\cal E}(-0,t) .
\ee
Hence the density of polaritons increases by a factor $c/v_{\rm g}$
when entering the medium, while there total number is conserved since
the pulse length is changed by the inverse factor. 

Boundary effects can also be used for a controlled deformation of a
stored light pulse. When the polariton reaches the back end of the
medium before $\cos\theta(t)$ approaches unity, 
the pulse will be spatially modulated.


\section{Limitations of quantum state-transfer}


In the preceeding sections we have discussed the control of 
propagation of the photon wavepackets under the assumptions of an
adiabatic evolution and small probe-field amplitudes.
We furthermore neglected ground-state decoherence and motional
(Doppler) effects. In the present section we will discuss the
validity of these approximations in detail.


\subsection{non-adiabatic corrections}


Let us first discuss the effect of non-adiabatic transitions 
i.e. let us take into account terms in first order of $\varepsilon$
in eqs.(\ref{Psi-full}) and (\ref{Phi-full}). 
For the following discussion it is sufficient to consider the semiclassical
limit, where the operator character of the variables as well as the Langevin
noise $F_{ba}$ can be disregarded. 
Up to $\varepsilon^1$ one obtains
\be
\Phi\approx \sin^2\theta(t) \frac{\gamma}{g^2 N} \, \dot\theta\, \Psi+
\frac{\sin^3\theta(t)}{\cos\theta(t)} \frac{\gamma}{g^2 N}
\dot\Psi.
\ee
On the same level of approximations we can replace $\dot\Psi$ on the right
hand side by $-c \cos^2\theta\partial\Psi/\partial z$. This gives
\be 
\Phi\approx \sin^2\theta(t) \frac{\gamma}{g^2 N} \, \dot\theta\, \Psi-
\sin^3\theta(t)\cos\theta(t) \frac{\gamma}{g^2 N} c\frac{\partial}{\partial z}
\Psi.
\ee
Substituting this into eq.(\ref{Psi-full}) yields the lowest-order
non-adiabatic corrections to the propagation equation of the dark-state
polariton:
\be
&&\left[\frac{\partial}{\partial t}+c\cos^2\theta(t)
\frac{\partial}{\partial z}\right]\Psi(z,t) =  -A(t)\Psi
+B(t)c\frac{\partial}{\partial z}\Psi +\nonumber\\
&&\quad\qquad
+\, C(t)c^2  \frac{\partial^2}{\partial z^2}\Psi
- D(t)c^3 \frac{\partial^3}{\partial z^3}\Psi,
\ee
where
\be
A(t) &=& \Bigl(\gamma+\frac{1}{2}\frac{\partial}{\partial t}\Bigr)
\left(\frac{\dot\theta^2\sin^2\theta}{g^2 N}\right),\\
B(t)&=& \frac{\sin\theta}{3 g^2 N} \frac{\partial^2}{\partial t^2} \sin^3\theta,\\
C(t) &=&  \Bigl(\gamma+\frac{1}{2}\frac{\partial}{\partial t}\Bigr)
\frac{\sin^4\theta\cos^2\theta}{g^2 N},\\
D(t) &=& \frac{\sin^4\theta\cos^4\theta}{g^2 N}.
\ee
$A(t)$ describes homogeneous losses due to 
non-adiabatic transitions followed by spontaneous emission.
$B(t)$ gives rise to a correction
of the polariton propagation velocity, $C(t)$ results in a pulse spreading 
by dissipation of high spatial-frequency components and $D(t)$ leads to
a deformation of the polariton. 

\begin{figure}[ht]
\centerline{\epsfig{file=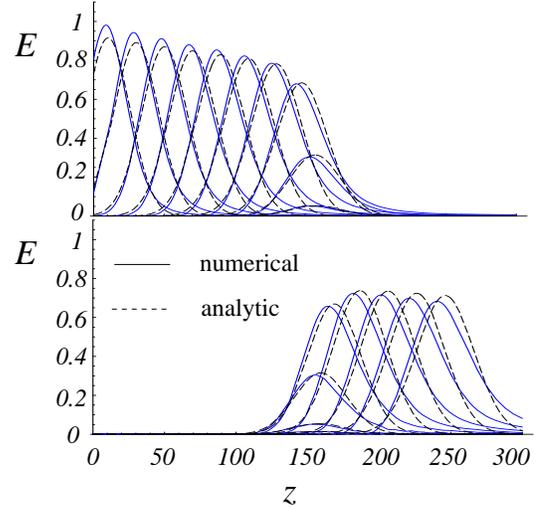,width=7 cm}}
\vspace*{2ex}
\caption{Deceleration (top) and re-aceleration (bottom) of a
Gaussian light pulse $\exp[-(t-500)^2/(200)^2]$ with
$\cot\theta(t)=0.363\bigl(1-(2/\pi)\, {\rm arccot}[5
 (1-0.5 \tanh[0.005(t-2000)]$
$+0.5 \tanh[0.005*(t-3200)])]\bigr)$.
The parameters are  $g^2 N/\gamma^2=2.5, \alpha=625, c=1$.
Pulses are shown for $t=$600, 800, 1000, 1200, 1400, 1600, 1800, 2000,
2200, and 2400 in the top figure and for $t=$2800, 3000, 3200, 3400, 3800, 4000
in the bottom figure. One recognizes good
agreement between numerical solution of the propagation equations
(solid line) and analytic approximation (dashed line).}
\label{stop-nonadiabat}
\end{figure}

Since all coefficients depend only on time, the propagation equation can
be solved by a Fourier transform in space $\Psi(z,t)=\int {\rm d}k\, 
\widetilde\Psi(k,t)\, {\rm e}^{-ikz}$. This yields
\be
\widetilde\Psi(k,t)&=&\widetilde\Psi(k,0)\, \exp\left\{i k
\int_0^t\!\!\!{\rm d}t^\prime\Bigl[v_{\rm gr}(t')+c B(t')\Bigr] \right\}
\times\nonumber\\
&&\times\exp\left\{-ik^3 c^3 \int_0^t\!\!\!{\rm d}t^\prime D(t')\right\}
\times\label{analytic-solution}\\
&&\times
\exp\left\{-\int_0^t\!\!\!{\rm d}t^\prime 
\Bigl[A(t')+k^2 c^2 C(t')\Bigr]\right\}.
\nonumber
\ee
The last term contains all losses due to non-adiabatic corrections. 
In order to
neglect dissipation, the integral in the exponent of this term
needs to be small compared to unity. Taking into account
$\dot\theta(0)=\dot\theta(\infty)=0$ this results in the two conditions
\be
\gamma k^2 c^2 \int_0^\infty\!\!{\rm d}t\,
\frac{\sin^4\theta(t)\cos^2\theta(t)}{g^2N}\ll 1,
\ee
and 
\be
\gamma\int_0^\infty\!\!\!{\rm d}t\, \frac{\dot\theta^2\sin^2\theta}{g^2 N}
=\gamma\int_0^\infty\!\!\!{\rm d}t\, \frac{\dot\theta^2}{g^2 N+\Omega^2(t)}
\ll 1.
\label{adb}
\ee
The first condition can be brought into
a transparent form when replacing $\sin^4\theta$ by unity, which results in
\be
\gamma k^2 c^2 \int_0^\infty\!\!{\rm d}t\, \frac{\cos^2\theta(t)}{g^2 N}
=\frac{\gamma k^2 c z}{g^2 N}\ll 1
\ee
for all relevant $k$. 
Here $z=c\int_0^\infty {\rm d}t\, \cos^2\theta(t)$ 
is the propagation depth of the
polariton inside the medium, and we  have assumed 
$\cos\theta(0)=1$ and $\cos\theta(\infty)=0$ for simplicity.
Noting that the range of relevant spatial
Fourier frequencies is determined 
by the inverse of the initial pulse length $L_p$,
i.e. setting $k\sim L_p^{-1}$, 
one finds 
\be
z\ll \frac{g^2 N}{\gamma c}\, L_p^2\quad{\rm or}\quad z\ll \sqrt{\alpha} L_p,
\label{z-limit}
\ee
where $\alpha = g^2 N z/\gamma c$ is again the opacity of the medium 
without EIT. 
To illustrate this limitation we have shown in Fig.\ref{stop-nonadiabat}
the slow-down and successive acceleration of a light pulse obtained from a numerical
solution of the full 1-d propagation problem as well as the analytical
approximation following from eq.(\ref{analytic-solution}). Here 
$\frac{g^2 N}{\gamma c} \frac{L_p^2}{z_{\rm max}} = 2$\, ($z_{\rm max}=250$). One clearly
recognizes a decay of the elm. excitation $E$ long before the deceleration and the
associated transfer to the atom systems sets in. The polariton energy
$I_\Psi\equiv \int {\rm d}z |\Psi(z,t)|^2$ decreases approximately exponentially
with the propagation distance $\Delta z$ of the pulse center. In the example 
$I_\Psi$ decays after $\Delta z=250$ approximately by a factor ${\rm e}^{-0.322}\approx 0.724$.


Since the spatial pulse length in the medium is related to
the initial group velocity $v_g^0$  (at time $t=0$ before the trapping) 
and to the initial pulse bandwidth 
as $L_p \sim v_g^0/\Delta \omega_p(0) $ and since 
$\Delta \omega_{tr}(0) =  \sqrt{\alpha}/\tau_d \approx\sqrt{\alpha} v_g^0/z$, 
for $v_g^0\ll c$, 
equation (\ref{z-limit}) is simply another way to state:
\be
\Delta \omega_{p} \ll \Delta \omega_{tr}(0). 
\ee
Hence the present analysis confirms the qualitative 
conditions for adiabatic following: 
the initial probe spectrum should be contained within the
original transparency window. Once again, this can be satisfied
if the medium is optically dense, $\alpha \gg 1$.

There is also a second condition (\ref{adb}) which is  
well-known from adiabatic  passage requirement
\cite{Fl-Manka96,Vitanov} and sets a limit to the 
rotation velocity $\dot\theta$ of the mixing angle and hence to
the deceleration/acceleration of the polariton. Introducing a
characteristic time scale of the  acceleration/deceleration period $T$
we obtain 
\be
T > {l_{abs} \over c} {v_g^0 \over c},
\ee
where $l_{abs} = c \gamma/g^2 N$ is the absorption length 
in the absence of EIT.
We note that for realistic experimental parameters the quantity 
on the right hand  side is extremely small on all relevant time scales. 
Hence, in practice the rate of change of the mixing angle does not 
significantly affect the adiabatic dynamics.


\subsection{weak field approximation}


The analysis of sec.IV also involves a perturbation expansion, valid 
when the control field is much stronger than the probe field. 
It is easy to see 
that this expansion is justified even when the
control-field  Rabi-frequency is reduced 
to zero. Making use of (\ref{s_cb_ad}) one finds:
$g^2 \langle{\hat E}^+{\hat E}\rangle/ |\Omega|^2 = 
\langle{\hat \sigma}_{cb} {\hat \sigma}_{bc}\rangle = \langle \hat\sigma_{cc}
\rangle$.
I.e., the ratio of the average intensities of quantum and 
control field
is proportional to the probability to find an atom
in state $c$. 
If the initial number-density 
of photons in the quantum field is much less
than the number-density
of atoms,  $\langle\hat\sigma_{cc}\rangle$ is always much smaller than 
unity. Therefore the mean  intensity of the quantum field 
remains small compared to that of the control field even when the latter 
is turned to zero. 


\subsection{decay of Raman coherence}


In the ideal scenario considered in sec.IV
we disregarded all processes 
resulting in a decay of coherences between the metastable states
$b$ and $c$. If such a decay at a rate $\gamma_0$ is taken into account,
the group velocity in the EIT medium cannot reach zero.
Nevertheless an (almost) complete transfer of the quantum state of
a propagating light pulse to a stationary matter excitation is possible.
Introducing phenomenogical decay into the 
equation of the Raman coherence ${\hat\sigma}_{bc}$ results in 
a modification of equation (\ref{bc}) according to 
\begin{eqnarray}
&&\ws_{ba}=-\frac{ i}{\Omega^*}  {\rm e}^{i\Delta k z} \left(
\frac{\partial}{\partial t}
\ws_{bc} + \gamma_0 \ws_{bc} \right) + F_{bc}. 
\end{eqnarray}
For formal consistency a Langevin noise term
associated with the decoherence needed to be introduced as well.  Such a
dephasing trivially modifies the propagation law of the polariton 
\begin{eqnarray}
\hat\Psi(z,t)&=& 
\exp\Bigl(-\gamma_0 \int_0^t\!\! d\tau \sin^2\theta(\tau)\Bigr)\times
\nonumber\\
&&\times \hat \Psi\biggl(z- c\int^t_0\!\!\!{\rm d}\tau
\cos^2\theta(\tau),0\biggr) + {\hat F}_{\Psi}.
\label{sol10}
\end{eqnarray}
As expected dephasing simply results in a decay of the polartion 
amplitude $\langle\hat\Psi\rangle$ but does not prevent light 
stopping.
A more detailed discussion of various realistic decoherence 
mechanisms and their influence on the fidelity of the
quantum memory will be presented 
in a subsequent publication.


\subsection{atomic motion}


We have shown above that adiabatically varying the
envelope  of the control field $\Omega$ in {\it time} a light pulse
of the probe field can be brought to a full stop. A spatially homogeneous 
variation of the group velocity can be realized, for instance, 
in the case when  the control and probe field are propagating
in orthogonal directions. However, in such a case the
spin-wave component of the polariton has a phase, which is
rapidly oscillating in $z$, since then $\Delta k=k_d-k_d^\parallel
=k_d(1-\cos\theta)=k_d$. For an atom at position $z_j$ one would have
\be
\hat \sigma_{bc}^j(t) =
|\hat \sigma_{bc}^j(t)|\, \exp\left\{i \frac{\omega_{ac}}{c}z_j\right\} 
 \exp\left\{-i \frac{\omega_{bc}}{c}z_j\right\},
\ee
where a transformation from the rotating frame back into
the lab frame was applied. 
While the second term corresponds to spatial oscillations with the 
small beat-frequency between pump and drive, the first term 
oscillates with an optical frequency. As a consequence the 
the polariton state is highly sensitive to variations in the atomic positions,
or in other words would dephase rapidly due to atomic motion. 
To retain a high fidelity of the quantum memory it would be
necessary to confine the motion of the atoms during the storage time 
to within a fraction of an optical wavelength, which is a very
stringent condition. 

To avoid this problem nearly co-propagating beams can be used,
such that $\Delta k\ll k_d$. In this case almost no photonic momentum is 
transferred to the atoms and the two-photon transition frequency would
experience 
only a very small Doppler-shift. Such a  configuration 
is therefore 
robust with respect to atomic motion, as 
electronic and motional degrees of freedom are completely decoupled. 

In a Doppler-free configuration propagation effects of the
control field need to be considered. In the case when the probe field
is weak compared to the control field at all times, the latter
propagates almost as in free space and  $\Omega(z,t)=\Omega(t-z/c)$. Due to the
resulting $z$-dependence of the group velocity, the trapping mechanism does
not exactly preserve the shape of the photon wavepacket. However, when the
probe pulse enters the medium with a group velocity much smaller than
the velocity of the control field  $v_g^0\ll c$,
the retardation of the control-field amplitude can be ignored
and $\Omega(t-z/c)\approx \Omega(t)$. This is a 
good approximation
as long as the time variation of the control field is sufficiently slow such 
that its spatial variation across the probe pulse length is small. 
This implies that the characteristic time of change of the control field 
should obey  ${T} \gg L_p/c$,  which is again
easily satisfied.

Before concluding we note that even in the case when no photon momentum
is transfered to atoms and therefore atomic motion does not result in 
altering the phase of the spin coherence, in practice it is still 
desirable to suppress the motion. This is because 
the light beams normally have a finite
cross-section and the atomic coherence is localized in the longitudinal 
direction and motion would tend to spread the localized excitation over the 
entire volume occupied by the gas. In this case the information about 
the pulse shape and the mode function will be lost after a sufficiently 
long storage interval. That is the reason why techniques involving 
cold trapped atoms  and buffer gas were used to effectively slow the atomic 
drifts  in recent 
experiments \cite{lui,phillips}.


\section{summary}


In conclusion we introduced the basic idea of quantum memory for light
based on dark-state polaritons and discussed their properties.
We have considered the influence of the main realistic imperfections on
trapping and re-acceleration and have shown that the technique is extremely 
robust. In particular, 
we demonstrated that an essential mechanism that enables the efficient quantum 
memory operation - the adiabatic following in polaritons - can take place
despite of transparency window narrowing. Subsequent papers 
will discuss the influence of other decoherence mechanisms on 
photon trapping and retrieval and will present the results of 
detail numerical
simulations of the trapping procedure.


\section*{acknowledgment}


The authors thank S.E. Harris, A. Imamoglu, C. Mewes, D. Phillips, 
M.O. Scully, R. Walsworth and  S. Yelin  for many stimulating 
discussions. This work was supported by the National Science Foundation via 
the grant to ITAMP and by the Deutsche Forschungsgemeinschaft.


\def\etal{\textit{et al.}}

\end{document}